# Mechanically-adjustable and electrically-gated single-molecule transistors

A. R. Champagne, A. N. Pasupathy, and D. C. Ralph

Laboratory of Atomic and Solid State Physics, Cornell University, Ithaca, NY, 14853.

We demonstrate a device geometry for single-molecule electronics experiments that combines both the ability to adjust the spacing between the electrodes mechanically and the ability to shift the energy levels in the molecule using a gate electrode.  With the independent *in-situ* variations of molecular properties provided by these two experimental "knobs", we are able to achieve a much more detailed characterization of electron transport through the molecule than is possible with either technique separately.  We illustrate the devices' performance using $C_{60}$ molecules.

A primary challenge in the field of single-molecule electronics [1,2] is to develop adjustable devices that can enable well-controlled, systematic experiments. If one uses techniques that measure only a current-voltage (*I-V*) curve, it can be difficult to determine even whether a molecule is present between electrodes, because nonlinear transport across tunnel junctions or metallic shorts can easily be mistaken for molecular signals [3]. Previous efforts to overcome this difficulty have employed two separate strategies for systematically adjusting a molecular device *in situ* to make changes that can be compared with theory. Electrostatic gating permits control of electron transport through a molecule by shifting its energy levels [4-7]. Mechanical adjustability, using scanning probes [8-11] or mechanically-controlled break junctions [12-15], enables manipulation of the device structure and the strength of bonding to electrodes. Here we report the implementation of both electrostatic gating and mechanical adjustability within the same single-molecule device. This combined capability enables a detailed characterization of electrical transport in molecules, providing understanding that is not possible with just gating or mechanical adjustability separately.

Our scheme for combining electrostatic gating and mechanical adjustability is to add a gate electrode to a mechanically-controlled break-junction (MCBJ). An MCBJ [16–19] consists of a narrow bridge of metal suspended above a flexible substrate (Fig. 1(a,b)). By bending the substrate, one can break the bridge and then adjust the spacing between the resulting electrodes. The main challenge in fabricating an electrically-gated MCBJ is to minimize the molecule-gate spacing to enable useful gating. We use as our gate electrode a degenerately-doped Si substrate, which allows us to employ standard lithographic techniques to produce a molecule-gate spacing as small as 40 nm, compared to ~1 μm substrate spacings achieved previously with MCBJs on rougher metal or glass substrates [18].

To fabricate the devices, we first grow a 250-nm thick $SiO_2$ film on top of a 200 μm-thick degenerately-doped silicon wafer. Using photolithography and a hydrofluoric-acid etch, we open windows in this thick oxide film and grow a thinner 40-nm oxide for the device regions. We then use photolithography, electron-beam lithography and liftoff to pattern Au lines 32 nm thick and 500 nm long, with a 50-nm wide constriction in the middle, connected to larger-area contact pads. The Au lines are positioned within the thin-oxide windows while the contact pads lie on the thicker-oxide regions. A timed buffered-hydrofluoric-acid etch is used to remove the $SiO_2$ from under the Au bridge, suspending it above the silicon substrate. A scanning electron microscope (SEM) image of a device before our Au bridge is broken is shown in Fig. 1(b).

We bend the substrate by placing a 15-mm long by 6-mm wide Si chip against two supports spaced 10 mm apart, and applying a pushing screw to the middle of the chip (Fig. 1(a)). The fine-threaded screw (1/80 inch pitch) is driven by a stepper motor via a series of reducing gears (factor of 100 reduction). Each chip contains 36 devices, of which 16 can be wired bonded at once. The amount of bending allowed by the Si chip (~ 0.3 mm over 10 mm) is generally not sufficient to break the metal bridge by mechanical motion alone. Therefore we first break the wires partially or fully at 4.2 K using electromigration, a technique used previously to make nm-scale gaps for molecular transistors [20,4-7]. Electromigration is accomplished by ramping a voltage across the Au wire until the resistance increases. If the electromigration process is stopped when the sample conductance reaches a few times $e^2/h$, mechanical motion can be used to complete the process of breaking the wire, resulting in the stepwise reduction in conductance as observed in other MCBJ devices [19] (inset of Fig. 1(c)). Electromigration can also be used to break the wires completely, giving typically decice resistances ranging from 100s of kΩ to 100s of MΩ.

To calibrate the motion, we performed measurements at 4.2 K on bare Au electrodes after electromigration was used to fully break the wires. The junction resistance increases smoothly upon bending the substrate (Fig. 1(c)). The resistance of a tunnel junction is expected to depend on the width $x$ of the tunneling barrier approximately as [21]

$$R \propto e^{2\kappa x}, \quad \kappa = \frac{\sqrt{2m_e\phi}}{\hbar},$$ (1)

where $\phi = 5.1$ eV is the Au work function and $m_e$ the electron mass. From the data in Fig. 1(c), we extract a calibration of $5.57 \pm 0.06$ pm per full rotation of the stepper motor. The average over our devices is $5.4 \pm 0.3$ pm/turn. The full range of motion available before the substrate breaks is generally 5 Å. A few devices (not included in the average) exhibited much less motion per turn, together with large gate leakage currents (> 1 nA at a gate voltage $V_g = 1$ V). In SEM images, these were identified as collapsed bridges.

If the substrate bends uniformly, the expected source-drain displacement is [19]

$$\delta x = \frac{3ut}{L^2} \delta y,$$ (2)

where $u$ is the length of the suspended bridge (500 nm), $t$ is the wafer thickness (200 μm), $L$ is the distance between the two support posts used for bending (10 mm), and $\delta y$ is the pushing screw displacement (3.1 μm per full rotation of the stepper motor). This formula predicts $\delta x = 9$ pm per full rotation, within a factor of 2 of the calibration result. The difference is similar to results for other MCBJs [16-19], and can be ascribed to uncertainties in the tunnel barrier height or to non-uniform bending of the substrate [19,22].

To characterize the performance of the gate electrode, we must insert into the device a molecule with low-lying energy levels, which can be shifted by $V_g$ to modulate current flow. We chose $C_{60}$ molecules because they have been used successfully to make single-molecule transistors [4,7]. To fabricate $C_{60}$ devices, we first clean unbroken Au wires in acetone, isopropanol, and oxygen plasma. We deposit a 0.2 mM solution of $C_{60}$ in toluene, and blow dry after 30 s. Then we cool to 4.2 K and perform electromigration until the device's conductance falls below a quantum of conductance  In approximately 30% of samples, one or more $C_{60}$ molecules bridge between the electrodes [4]. We identify these samples as the ones whose *I-V* curves display Coulomb-blockade characteristics -- with non-negligible current only for $|V|$ greater than threshold values that depend on the gate voltage $V_g$. Control junctions formed from bare Au electrodes did not exhibit such Coulomb-blockade characteristics. All the measurements that we will describe were conducted at 4.2 K.

The effect of $V_g$ and $V$ on conductance, at fixed source-drain displacement, is shown in Fig. 2 for three samples at two different source-drain distances each. Fig. 3(a) shows the corresponding Coulomb-blockade *I-V* curves for sample #1 at $V_g = 2.5$ V. The dark regions on the left and/or right of each panel in Fig. 2 correspond to low-current regions of Coulomb blockade. Bright regions denote large *dI/dV*, where the applied source-drain voltage provides sufficient energy for electrons to tunnel via the molecule and initiate current flow. The threshold *V* required for current flow depends on $V_g$, which shifts the energy of the molecular states with respect to the Fermi energy of the electrodes. The energy to add an electron to a molecular level can be tuned to zero for a particular value of gate voltage $V_g = V_{deg}$.

As can be seen in Fig. 2, the sensitivity of the transport characteristics to $V_g$ varies between samples. This is to be expected because the gate-molecule capacitance $C_g$ will depend on the position of the molecule within the junction. The positive and negative slopes of the tunneling thresholds in Fig. 2 correspond, respectively, to the ratios $C_g / (C_g + C_d) \approx C_g / C_d$ and $C_g / C_s$, where $C_s$ and $C_d$ are the molecule-source and molecule-drain capacitances. From the measured slopes in Fig. 2, we find ratios of $C_d / C_g$ and $C_s / C_g$ between 2200 and 57. Our smallest value of $(C_s + C_d)/C_g$ is 148 and the largest values are several thousands. The largest gate voltage that we can apply routinely is $\pm 12$ V, after which the devices fail due to oxide breakdown under the electrode pads. Given this range of $V_g$, we are able to shift the quantum states on a $C_{60}$ molecule by an energy as large as $\Delta E = e[C_g / (C_s + C_d + C_g)] \, \Delta V_g = 160$ meV.

The capability to manipulate the energy levels on the $C_{60}$ molecules with a gate allows a detailed analysis of the changes caused by mechanical displacement, providing understanding that is not achievable with either motion or gating alone. First, note in Fig. 2 that mechanical motion can change the slopes of the tunneling thresholds. This is a consequences of changes in $C_s$ and $C_d$. In Fig. 3(b) we plot $C_s/C_g$ and $C_d/C_g$ for samples #2 and #3. For sample #2, both capacitance ratios change with motion, but for sample #3 only the drain capacitance varies significantly between 0 and 0.8 Å displacement, indicating that the motion predominantly modifies the molecule-drain distance.

A second change caused by the mechanical motion is that the overall device conductance decreases with increasing displacement (note Fig. 3(a) and the color-scale changes in Fig. 2). In Fig. 3(c), we plot as a function of displacement the conductances for samples #2 and #3 at large enough $V$ that Coulomb-blockade effects are negligible. For sample #3, for which the capacitance ratios indicated that only the molecule-drain distance changes significantly with displacement, the conductance decreases exponentially with distance, with a decay length approximately equal to that for bare Au junctions. For sample #2, for which we found from the capacitances that both the source and drain junctions change with displacement, the conductance change is not a single exponential and the rate of change is slower than for bare Au. This is consistent with a situation in which the conductances of the two tunnel junctions are comparable and the motion is divided between both junctions. The information from gate-voltage measurements therefore provides an explanation for the different dependences of the conductance on electrode displacement.

A third type of change caused by the electrode displacement is perhaps less intuitive than the changes in capacitances and conductance -- moving the source and drain electrodes can shift the gate voltage corresponding to the degeneracy point in the Coulomb blockade curves (Fig. 2(c,d) and Fig. 3(d)). This effect demonstrates that the electrodes carry a charge even when $V = 0$, so that when they move they perform work and shift the energy of the molecular states relative to the Fermi energy of the leads. The tunneling threshold $V$ can be changed by as much as 25 mV by a displacement of 1.2 Å (Fig. 3(e)). Hence, these devices can exhibit transistor action controlled entirely by means of mechanical motion, as well as the ordinary electrostatic transistor action enabled by changing $V_g$. Related energy shifts have been seen in scanning-tunneling-microscope experiments on metal nanoparticles [23,24], where it was suggested that they were related to a contact potential $\Delta\varphi$ between the electrodes and the nanoparticle,

$$\Delta V_{\mathrm{deg}} = \Delta\varphi\left(\frac{\Delta C_s + \Delta C_d}{C_g}\right). \tag{3}$$

Our devices permit direct measurements of $(\Delta C_s + \Delta C_d)/C_g$, and we find that the measured values of $\Delta V_{\mathrm{deg}}$ would correspond to different values of $\Delta\varphi$: -7 ± 5 mV for sample #1, 64 ± 3 mV for sample #2, and -29 ± 5 mV for a fourth sample. This means that $\Delta V_{\mathrm{deg}}$ is not determined only by an intrinsic work-function difference, but may also be influenced, for instance, by charged defects.

We have checked whether an applied gate voltage produces unwanted mechanical motion.  For both bare Au junctions and $C_{60}$ devices with weak gate dependence, we have swept $V_g$ in the range ±12 V at fixed $V$ and found less than 10% change in current, comparable to fluctuations due to charge rearrangements as we sweep $V_g$.  Based on conductance versus displacement measurements on the same samples, this corresponds to an upper bound of 5 pm for $V_g$- induced displacements.

Our gated MCBJs provide the capability to make measurements on molecules with simultaneous tuning of both the spacing between electrodes (over a range of 5 Å) and the energy levels (by 160 meV on a $C_{60}$ molecule).  We find that mechanical motion changes a molecule's capacitances, conductance, and Coulomb-blockade-degeneracy gate voltage.  Gate-voltage control permits detailed and mutually consistent characterizations of all three effects.  We anticipate that this new experimental tool will enable systematic studies of fundamental questions such as how changes in molecular conformation affect electrical conduction, and how transport in molecules varies as one tunes between the weak-coupling Coulomb-blockade and strong-coupling Kondo regimes [5].

We thank Kirill Bolotin, Ferdinand Kuemmeth, and Sergey Kiselev for discussions and technical help. This work is supported by the NSF through the Cornell Center for Nanoscale Systems (EEC-0117770), DMR-0244713, and use of the Cornell Nanoscale Facility CNF/NNIN, and by ARO (DAAD19-01-1-0541).

**Figure captions**

FIG. 1.  (a) The design of the experimental apparatus.  (b) Scanning electron microscope image of a MCBJ sample with a silicon substrate gate, before breaking the gold bridge. The image was acquired at a $78^o$ tilt angle to reveal the 40 nm spacing between the gold bridge and the substrate. (Inset) A sample that has been broken by electromigration at 4.2 K and subsequently warmed to room temperature.  (c) Resistance of a bare Au tunnel junction versus source-drain displacement at 4.2 K, from which the amount of mechanical motion can be calibrated.  (Inset)  When an atomic-scale metal contact remains after electromigration, mechanical motion can be used to complete the process of breaking the wire.

FIG. 2.  Color-scale plots of $dI/dV$ versus $V$ and $V_g$ for three $C_{60}$ gated MCBJ devices at 4.2 K, each at two positions of source-drain displacement.  $x_0$ represents the initial displacement. The relative displacements were determined using the bare-Au calibration value, 5.4 pm/turn.  (a-b) Sample #1.  (c-d) Sample #2.  (e-f) Sample #3. The color scale is logarithmic in (c,d) and linear in (a,b,e&f).  Scans for samples #1, #2 and #3 were acquired every 0.16 Å, 0.27 Å and 0.11 Å, respectively, and showed a smooth evolution with motion.

FIG. 3.  (a) Current-voltage curves for sample #1 at different source-drain displacements, $V_g = 2.5$ V.  (b) $C_s/C_g$ (open symbols) and $C_d/C_g$ (filled symbols) as a function of source-drain displacement, for two samples.  For sample #2, use right and top axes, and for sample #3, use left and bottom axes.  (c) Sample conductance $G = dI/dV$ versus source-drain displacement at large enough biases ($V = 235$ mV for sample #2, 150 mV for sample #3) that Coulomb-blockade effects are not significant.  (d) $V_{deg}$ as a function of mechanical

motion for sample #2. (e) Color-scale plot of $dI/dV$ versus $V$ and source-drain displacement for sample #3, with $V_g = 0$.

## References

1. A. Nitzan and M. A. Ratner, *Science* **300**, 1384 (2003).

2. J. R. Heath and M. A. Ratner, *Physics Today* **56**, 43 (May, 2003).

3. C. N. Lau et al., *Nano Lett.* **4**, 569 (2004).

4. H. Park et al., *Nature* **407**, 57 (2000).

5. J. Park et al., *Nature* **417**, 722 (2002).

6. W. J. Liang, et al., *Nature* **417**, 725 (2002).

7. L. H. Yu and D. Natelson, *Nano Lett.* **4**, 79 (2004).

8. L. A. Bumm et al., *Science* **271**, 1705 (1996).

9. D. Porath et al., *Phys. Rev. B* **56**, 9829 (1997).

10. B. C. Stipe, M. A. Rezaei, and W. Ho, *Science* **280**, 1732 (1998).

11. D. Cui et al., *Science* **294**, 571 (2001).

12. M. A. Reed et al., *Science* **278**, 252 (1997).

13. C. Kergueris et al., *Phys. Rev. B* **59**, 12505 (1999).

14. R. H. M. Smit et al., *Nature* **419**, 906 (2002).

15. B. Q. Xu, X. Y. Xiao, and N. J. Tao, *J. Am. Chem. Soc.* **125**, 16164 (2003).

16. J. Moreland and J. W. Ekin, *J. Appl. Phys.* **58**, 3888 (1985).

17. C. Zhou et al., *Appl. Phys. Lett.* **67**, 1160 (1995).

18. J. M. van Ruitenbeek et al., *Rev. Sci. Instrum.* **67**, 108 (1996).

19. N. Agrait, A. L. Yeyati, and J. M. van Ruitenbeek, *Phys. Rep.* **377**, 81 (2003).

20. H. Park et al., *Appl. Phys. Lett.* **75**, 301 (1999).

21. R. Wiesendanger, *Scanning Probe Microscopy and Spectroscopy* (Cambridge University Press, 1994).

22. B. Ludoph, Ph.D. Thesis. (Leiden University, 1999).

23. A. E. Hanna and M. Tinkham, *Phys. Rev. B* **44**, 5919 (1991).

24. Z. Y. Rong, A. Chang, L. F. Cohen and E. L. Wolf, *Ultramicroscopy* **42**, 333 (1992).

**(a)**

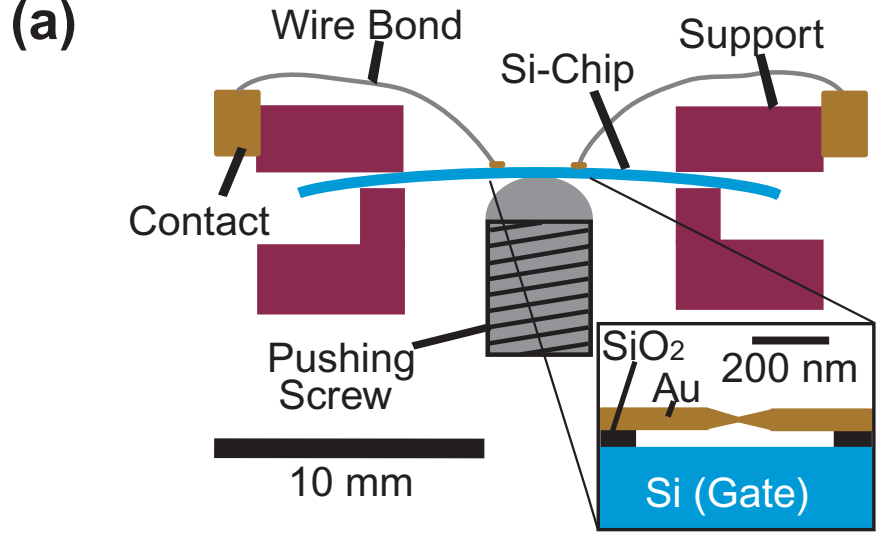

Wire Bond · Si-Chip · Support · Contact · Pushing Screw · SiO₂ · Au · 200 nm · Si (Gate) · 10 mm

**(b)**

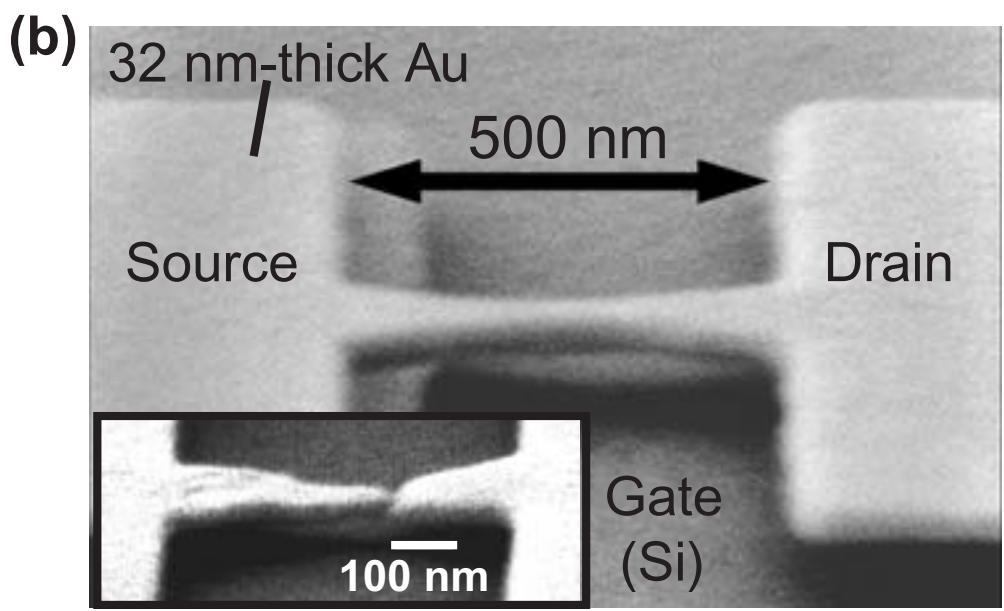

32 nm-thick Au · 500 nm · Source · Drain · Gate (Si) · 100 nm

**(c)**

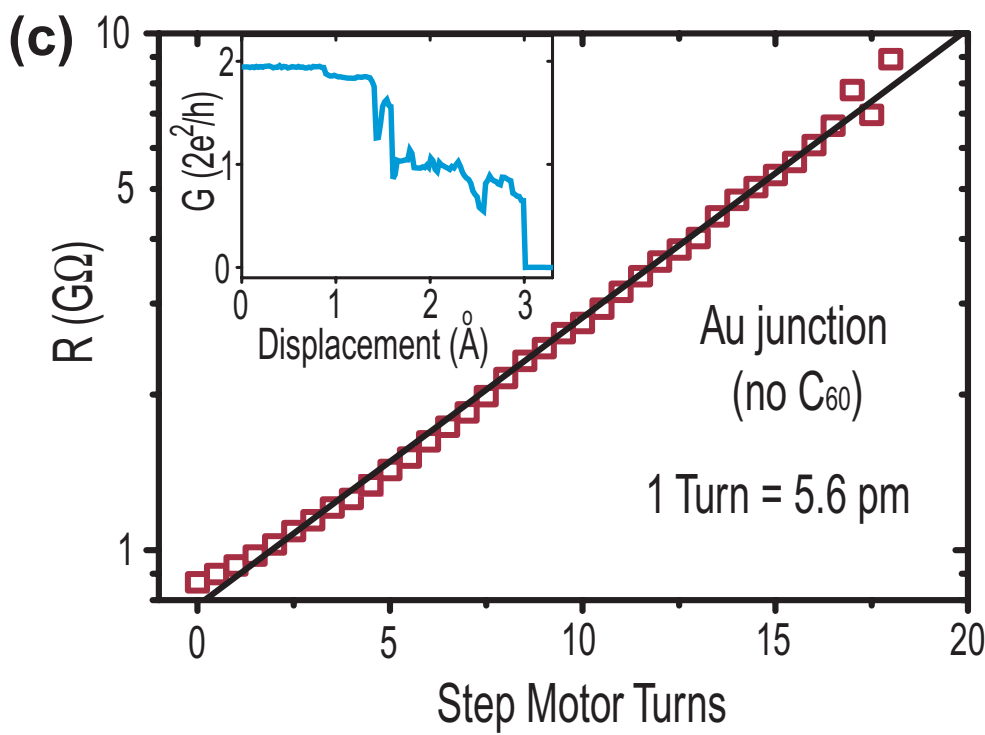

R (GΩ) vs Step Motor Turns · G (2e²/h) vs Displacement (Å) · Au junction (no C₆₀) · 1 Turn = 5.6 pm

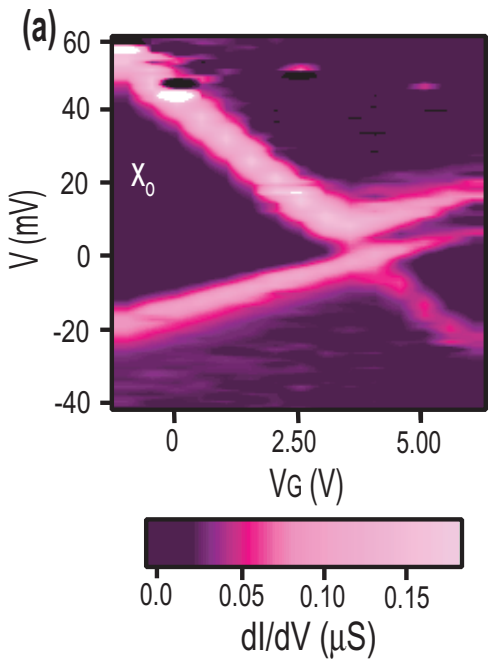

**(a)**

$X_0$

V (mV): 60, 40, 20, 0, -20, -40

$V_G$ (V): 0, 2.50, 5.00

0.0  0.05  0.10  0.15
dI/dV (µS)

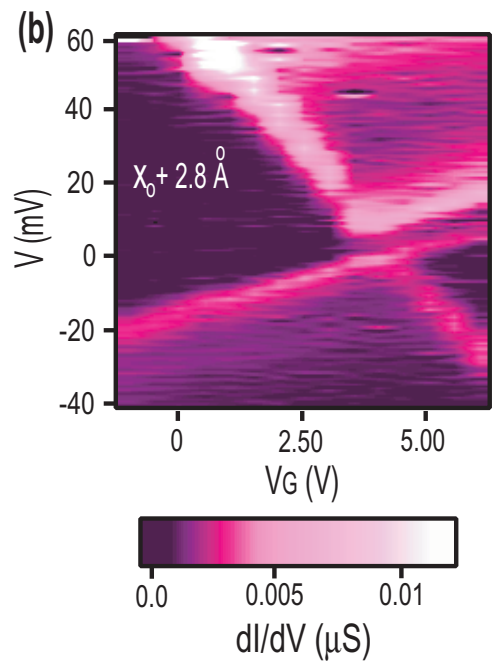

**(b)**

$X_0$+ 2.8 Å

V (mV): 60, 40, 20, 0, -20, -40

$V_G$ (V): 0, 2.5, 5.0

0.0  0.005  0.01
dI/dV (µS)

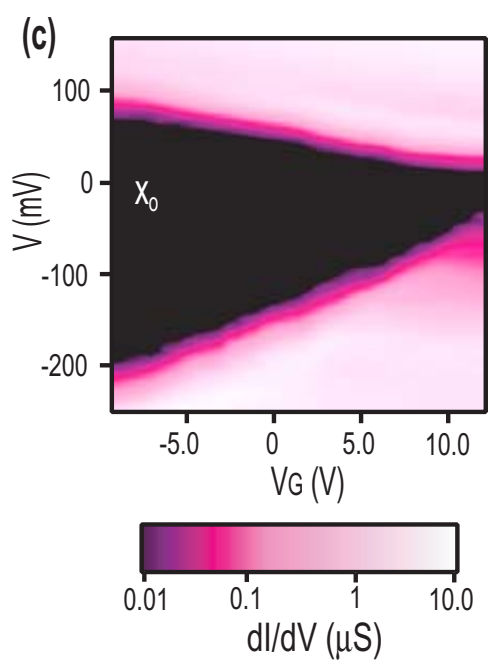

**(c)**

$X_0$

V (mV): 100, 0, -100, -200

$V_G$ (V): -5.0, 0.0, 5.0, 10.0

0.01  0.1  1  10.0
dI/dV (µS)

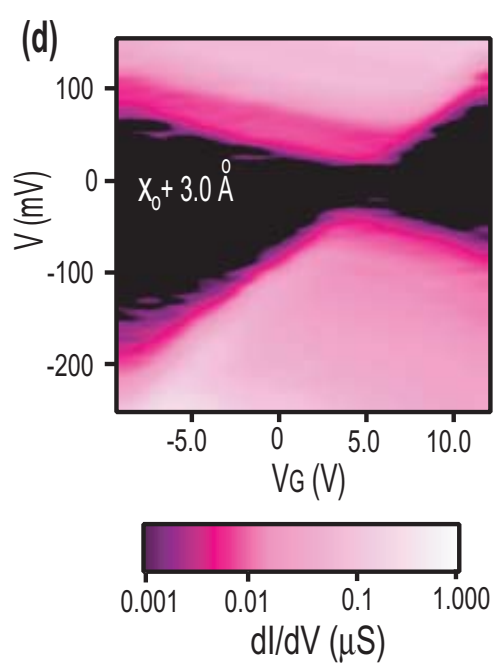

**(d)**

$X_0$+ 3.0 Å

V (mV): 100, 0, -100, -200

$V_G$ (V): -5.0, 0.0, 5.0, 10.0

0.001  0.01  0.1  1.000
dI/dV (µS)

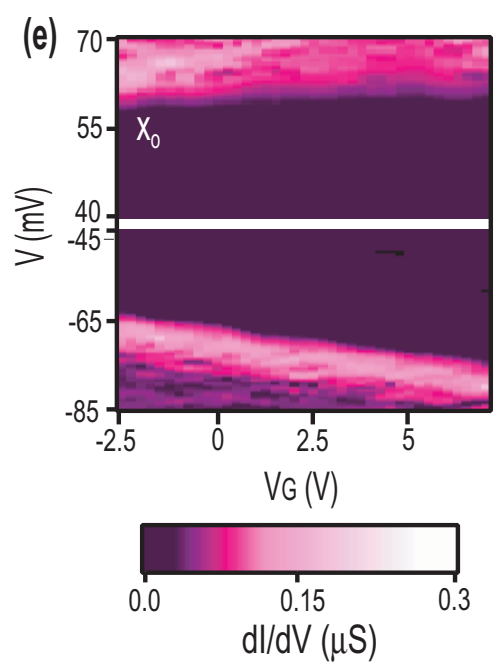

**(e)**

$X_0$

V (mV): 70, 55, 40, -45, -65, -85

$V_G$ (V): -2.5, 0, 2.5, 5

0.0  0.15  0.3
dI/dV (µS)

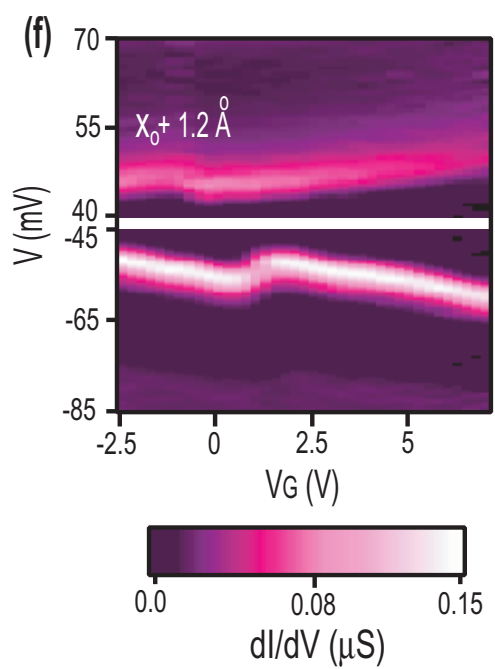

**(f)**

$X_0$+ 1.2 Å

V (mV): 70, 55, 40, -45, -65, -85

$V_G$ (V): -2.5, 0, 2.5, 5

0.0  0.08  0.15
dI/dV (µS)

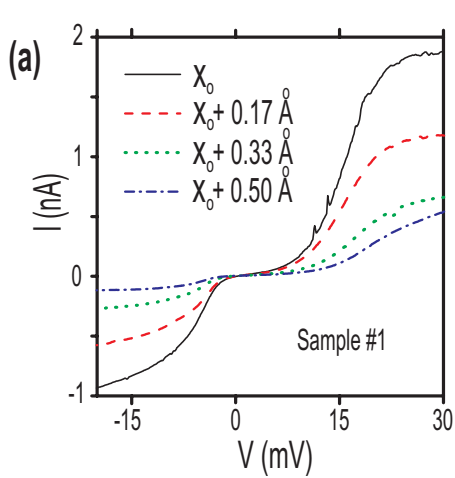

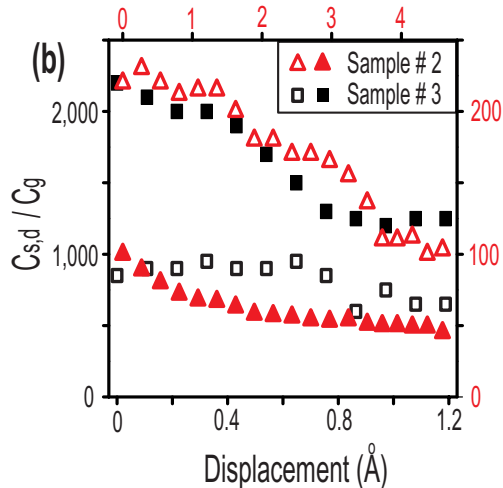

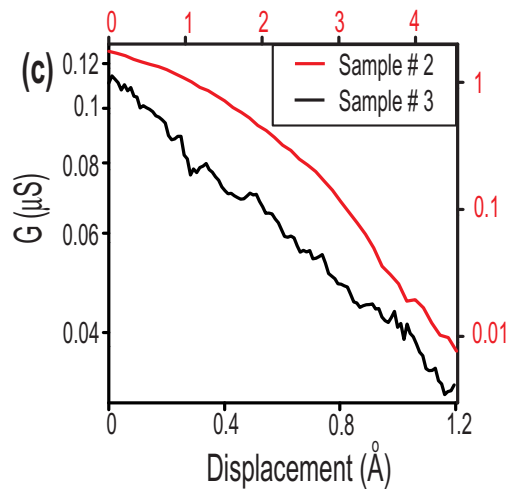

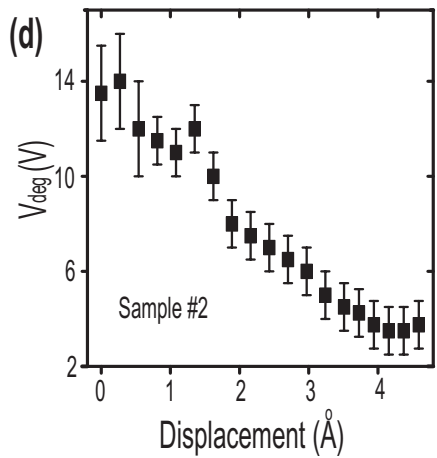

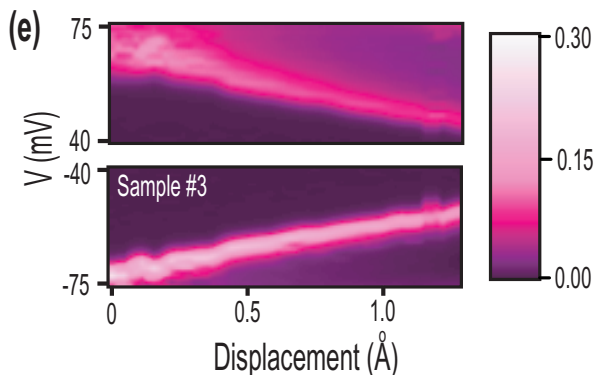